\documentclass[preprint,authoryear,12pt]{elsarticle}
\usepackage[left=1.7cm,right=2.8cm]{geometry}
\usepackage{amssymb,amsmath,amstext,psfrag}
\usepackage{accents}

\journal{International Journal of Plasticity}

\begin{document}

\begin{frontmatter}

\title{Analysis of some basic approaches to
finite strain elasto-plasticity in view of reference change}

\author{A. V. Shutov}
\author{J. Ihlemann}

\address{Technische Universit\"at Chemnitz, Department of Solid Mechanics, Chemnitz, Germany}

\begin{abstract}
There is a large
variety of concepts used to generalize the classical
Prandtl-Reuss relations of infinitesimal elasto-plasticity to finite
strains. In this work, some basic approaches are compared in a
qualitative way with respect to a certain invariance property.
These basic approaches include the additive hypoelasto-plasticity with
corotational stress rates,
additive plasticity in the logarithmic strain space, and
multiplicative hyperelasto-plasticity.

The notion of weak invariance is introduced in this study.
Roughly speaking, a material model is weakly invariant under a certain transformation of the
local reference configuration if this reference change
can be neutralized by a suitable transformation of initial conditions,
leaving the remaining constitutive relations intact.
We analyse the basic models in order to find out if
they are weakly invariant under arbitrary volume-preserving
transformations of the reference configuration.

It is shown that the weak invariance property
corresponds to a generalized symmetry which
provides insights into underlying
constitutive assumptions. This property
can be used for a systematic study of different frameworks
of finite strain elasto-plasticity. In particular, it
can be used as a classification criterion.
\end{abstract}


\end{frontmatter}


\section*{Nomenclature}
\begin{tabbing}
$\mathbf{F}$ \quad \quad \quad \quad \quad \quad    \= deformation gradient\\
$\mathbf{C}$ \>  right Cauchy-Green tensor \\
$\mathbf{B}$ \>  left Cauchy-Green tensor \\
$\mathbf{L}$ \>  velocity gradient tensor \\
$\mathbf{D}$ \>  strain rate tensor (stretching tensor) \\
$\mathbf{W}$         \>  continuum spin tensor (vorticity tensor) \\
$\mathbf{V}$         \>  left stretch tensor \\
$\mathbf{H}$         \>  Lagrangian logarithmic strain (Hencky strain) \\ 
$\mathbf{1}$               \>  idenity tensor \\
$\mathbf{T}$               \>  Cauchy stress tensor (true stresses) \\
$\mathbf{S}$               \>  Kirchhoff stress tensor \\
$\widetilde{\mathbf{T}}$    \>  2nd Piola-Kirchhoff stress tensor \\
$\hat{\mathbf S}$           \> weighted stress tensor operating on the
stress-free configuration \\
$\accentset{o}{\mathbf S}$  \>  objective stress rate \\
$\boldsymbol{\varepsilon}$       \>  infinitesimal strain tensor \\
$\boldsymbol{\sigma}$       \>  infinitesimal stress tensor or stress measure conjugate to $\mathbf{H}$ \\
$\mathbf{\Omega}$    \>  spin tensor \\
$\mathbf{R}$    \>  rotational part of $\mathbf{F}$ \\
$\dot{\mathbf{X}}$                \>  material time derivative \\
$\mathbf{X}^{\text{D}}$ \>  deviatoric part of a second-rank tensor \\
$\det(\mathbf{X})$               \>  determinant of a second-rank tensor \\
$\overline{\mathbf{X}}$ \>  unimodular part of a second-rank tensor \\
$\mathbf{X}^{\text{T}}$ \>  transposition of a second-rank tensor \\
$\text{sym}(\mathbf{X})$ \>  symmetric part of a second-rank tensor \\
$\text{skew}(\mathbf{X})$ \>  skew-symmetric part of a second-rank tensor \\
$\text{tr}(\mathbf{X})$        \>  trace of a second-rank tensor \\
$ \| \mathbf{X} \| $       \> Frobenius norm of a second-rank tensor \\
$ \mathbf{X} : \mathbf{Y} $  \> scalar product of two second-rank tensors \\
$t, t', t_0$              \>  time instances (typically $t_0 \leq t' \leq t$) \\
$\mathcal{Z}_0$       \>  initial state \\

\end{tabbing}

\section{Introduction}

The main purpose of the mathematical material theory is to provide some general
concepts useful in construction and analysis of
appropriate constitutive relations \citep{TruesdNoll, HauptKonzepte}.
In particular, any systematic study of competitive approaches
to finite strain elasto-plasticity should be based on a set of sound principles.
In this context, the following popular concepts and criteria have proved to be especially useful:
\begin{itemize}
\item the principle of material objectivity or material frame-indifference
\citep{Noll1958, TruesdNoll, Truesdell1969, Bressan1972, Svendsen1999, BertramSvendsen2001, Muschik2008, Korobeynikov2008},
\item requirement of thermodynamic consistency \citep{ColemanNoll1963, Coleman1964, Truesdell1969, Halphen1975},   
\item prevention of the shear oscillatory response  \citep{Lehmann1972, Dienes1979, Nagtegaal1982},
\item prevention of the energy dissipation within the elastic range \citep{SimoPister1984, BruhnsXiao1999}.
\end{itemize}

In this study we introduce the notion of weak invariance of constitutive relations under
arbitrary isochoric change of the local reference configuration.
This weak invariance cannot be attributed to general principles of material
modeling. Instead, we will demonstrate that it is a powerful tool of analysis and
classification, especially useful
within finite strain elasto-plasticity.

We start by recalling that
the local reference configuration is, in general, a fictitious configuration, set at will
to designate material points in a unique manner.
Hence, any arbitrary volume-preserving
change of the local reference configuration is always possible since
the transformed configuration can act as a new reference.
The main idea of the paper is as follows. The constitutive models of elasto-plasticity
can be subdivided into two major groups: those models which are
weakly invariant under arbitrary volume-preserving reference changes
and those which do not fulfill this invariance.

In order to get an intuitive insight into the weak invariance, we start with a simple example.
First, let us consider the well-known constitutive relations governing a elastic-perfectly-plastic material behaviour in the
geometrically linear setting. These relations include the classical Prandtl-Reuss kinematic equations
\begin{equation}\label{Prandtl-Reuss}
\boldsymbol{\varepsilon} = \boldsymbol{\varepsilon}_{\text{e}} + \boldsymbol{\varepsilon}_{\text{p}},
\end{equation}
where symmetric tensors $\boldsymbol{\varepsilon}$, $\boldsymbol{\varepsilon}_{\text{e}}$, $\boldsymbol{\varepsilon}_{\text{p}}$
stand for the total, elastic, and plastic strains, respectively.
The isotropic Hooke's law implies that the stress tensor $\boldsymbol{\sigma}$
is a linear function of $\boldsymbol{\varepsilon}_{\text{e}}$
\begin{equation}\label{IsotHooke}
\boldsymbol{\sigma} = k \ \text{tr}(\boldsymbol{\varepsilon}_{\text{e}}) \mathbf{1} +
2 \mu \boldsymbol{\varepsilon}^{\text{D}}_{\text{e}}.
\end{equation}
Here, $k > 0$, $\mu > 0$ are material parameters and
$( \cdot )^{\text{D}}$ represents the deviatoric part.
The Mises-Huber yield function with a constant
yield stress $K > 0$ takes the form
\begin{equation}\label{vMyield}
f := \| \boldsymbol{\sigma^{\text{D}}}\| - \sqrt{2/3} K,
\end{equation}
where $\| \mathbf{X} \| : = \sqrt{ \text{tr}(\mathbf{X} \mathbf{X}^{\text{T}})}$.
A six-dimensional flow rule
is introduced in combination with the Kuhn-Tucker constraints:
\begin{equation}\label{FlowRule}
\dot{\boldsymbol{\varepsilon}}_{\text{p}} =
\lambda_{\text{p}} \frac{\displaystyle \boldsymbol{\sigma^{\text{D}}}}{\displaystyle \| \boldsymbol{\sigma^{\text{D}}}\|},
\end{equation}
\begin{equation}\label{KuhnTuckerSmalSr}
f \leq 0, \  \lambda_{\text{p}} \geq 0, \ f \lambda_{\text{p}} =0.
\end{equation}
Here, $\lambda_{\text{p}}$ stands for the plastic multiplier.
Within the local approach, the history of the total strain tensor $\boldsymbol{\varepsilon}(t)$
is given. The system of equations is closed by appropriate initial conditions imposed on plastic strains
\begin{equation}\label{IniCond}
\boldsymbol{\varepsilon}_{\text{p}}|_{t=t_0} = \boldsymbol{\varepsilon}_{\text{p}}^0, \quad \text{tr} (\boldsymbol{\varepsilon}_{\text{p}}^0) =0.
\end{equation}
Note that the flow rule \eqref{FlowRule}
is incompressible, which is a typical assumption
in the context of metal plasticity. Thus, $\text{tr} (\boldsymbol{\varepsilon}_{\text{p}}) \equiv 0$.

Let us consider an arbitrary second-rank tensor
$\boldsymbol{\varepsilon}_0 = const$ such that $\text{tr} (\boldsymbol{\varepsilon}_0) =0$.
The system \eqref{Prandtl-Reuss}---\eqref{IniCond} has the following remarkable property.
If the prescribed loading programm is shifted according to
\begin{equation}\label{Shift}
\boldsymbol{\varepsilon}^{\text{new}}(t) := \boldsymbol{\varepsilon}(t) - \boldsymbol{\varepsilon}_{0},
\end{equation}
and the new initial conditions are obtained by the same shift
\begin{equation}\label{SameShift}
\boldsymbol{\varepsilon}^{\text{new}}_{\text{p}}|_{t=t_0} = \boldsymbol{\varepsilon}_{\text{p}}^0 - \boldsymbol{\varepsilon}_0,
\end{equation}
then the same stress response $\boldsymbol{\sigma}(t)$ is predicted by the constitutive euqations.

\textbf{Proof:}
In order to check this invariance property, we
consider the following candidate function for the new plastic strain
\begin{equation}\label{NewPlastStrain}
\boldsymbol{\varepsilon}^{\text{new}}_{\text{p}}(t) := \boldsymbol{\varepsilon}_{\text{p}}(t) - \boldsymbol{\varepsilon}_0.
\end{equation}
Let us show that this candidate function
corresponds to the new solution.
For this candidate, it follows immediately from \eqref{Shift} and \eqref{NewPlastStrain} that the elastic strains remain invariant
\begin{equation}\label{ElStrainInvar}
\boldsymbol{\varepsilon}^{\text{new}}_{\text{e}}(t) \stackrel{\eqref{Prandtl-Reuss}}{=}
\boldsymbol{\varepsilon}^{\text{new}}(t) - \boldsymbol{\varepsilon}^{\text{new}}_{\text{p}}(t) \stackrel{\eqref{Shift}, \eqref{NewPlastStrain}}{=}
\boldsymbol{\varepsilon}(t) -
\boldsymbol{\varepsilon}_{\text{p}}(t) \stackrel{\eqref{Prandtl-Reuss}}{=} \boldsymbol{\varepsilon}_{\text{e}}(t).
\end{equation}
Therefore, according to the
Hooke's law \eqref{IsotHooke}, the same stresses are predicted:
\begin{equation}\label{Samestresses}
\boldsymbol{\sigma}^{\text{new}}(t)=\boldsymbol{\sigma}(t), \quad f^{\text{new}}(t) = f(t).
\end{equation}
Thus, the Kuhn-Tucker conditions \eqref{KuhnTuckerSmalSr} are identically satisfied if we put
\begin{equation}\label{NewMult23}
\lambda^{\text{new}}_{\text{p}}(t) := \lambda_{\text{p}}(t).
\end{equation}
Moreover, the candidate function $\boldsymbol{\varepsilon}^{\text{new}}_{\text{p}}(t)$ satisfies
the initial condition \eqref{SameShift}.
By differentiating \eqref{NewPlastStrain}
with respect to time, it becomes obvious that $\boldsymbol{\varepsilon}^{\text{new}}_{\text{p}}(t)$
satisfies the evolution equation:
\begin{equation}\label{FormalDiffere}
\dot{\boldsymbol{\varepsilon}}^{\text{new}}_{\text{p}} \stackrel{\eqref{NewPlastStrain}}{=}
\dot{\boldsymbol{\varepsilon}}_{\text{p}} \stackrel{\eqref{FlowRule}}{=}
\lambda_{\text{p}} \frac{\displaystyle \boldsymbol{\sigma^{\text{D}}}}{\displaystyle \| \boldsymbol{\sigma^{\text{D}}}\|}
\stackrel{\eqref{Samestresses}, \eqref{NewMult23}}{=}
\lambda_{\text{p}}^{\text{new}} \frac{\displaystyle (\boldsymbol{\sigma}^{\text{new}})^{\text{D}}}{\displaystyle \| (\boldsymbol{\sigma}^{\text{new}})^{\text{D}}\|}.
\end{equation}
Therefore, the candidate $\boldsymbol{\varepsilon}^{\text{new}}_{\text{p}}(t)$ is indeed a solution
for the new local loading $\boldsymbol{\varepsilon}^{\text{new}}(t)$.
It remains to recall that $\boldsymbol{\sigma}^{\text{new}}(t)=\boldsymbol{\sigma}(t)$.
The assertion is proved
$\blacksquare$


In the remainder of this paper, the aforementioned invariance of the stress response under
the transformation \eqref{Shift} will
be seen as a \emph{weak} invariance. Here, the adjective ``weak" points to the fact that an additional freedom in transformation
of the initial conditions \eqref{SameShift} weakens the restrictions imposed on the
material model.\footnote{Interestingly, the transformation
\eqref{SameShift} was previously considered in combination
with the shift \eqref{Shift} by \cite{Rubin2001}
to question the status of the total strain as a state variable.}
In the next section, the notion of the weak invariance will
be reformulated in the finite strain context. As an extension to finite strains,
a weak invariance under isochoric change of the reference configuration will be introduced,
and a proper transformation of the initial conditions
(similar to \eqref{SameShift}) will be a crucial part of this concept.

Considering the generalization of small strain relations to finite strains,
the following aspect should be taken into account. If the small strain relations are
weakly invariant under the shift \eqref{Shift}, one
may expect that this invariance property will be inherited by the finite strain counterpart.

In the modern literature on elasto-plasticity there exists a large
variety of frameworks used to generalize the small strain
relations \eqref{Prandtl-Reuss}--\eqref{IniCond} to finite strains \citep{XiaoReviw}.
In this paper, some of the most popular approaches will be analysed from the viewpoint of the reference change.
These approaches include the additive hypoelasto-plasticity with
corotational stress rates (cf. Section 3), additive plasticity in the logarithmic strain space (cf. Section 4), and
multiplicative hyperelasto-plasticity (cf. Section 5).
In case of small strains and rotations, these approaches are reduced to the
geometrically linear theory \eqref{Prandtl-Reuss}--\eqref{IniCond}.
It is shown that some of these models are weakly invariant
under the reference change, but some of them not.
Thus, \emph{different models can be categorized using the weak invariance as a criterion}.
Moreover, the weak invariance corresponds to a certain generalized material symmetry.
Just like any other symmetry property, it brings new insights into our understanding
of the underlying constitutive assumptions.

\section{Weak invariance under the reference change}

Let $\mathbf{T}$ be the Cauchy stress tensor (also known as true stresses) at a certain material point and
$\mathbf{F}$ be the deformation gradient from the local reference configuration
$\widetilde{\mathcal{K}}$ to the current configuration $\mathcal{K}$.
The history of the deformation gradient from the initial state at $t_0$ up
to a certain time instance $t$ is captured by
$\mathbf{F}( t' ),  t' \in [t_0, t]$. Suppose that the state
of the material at $t_0$ is
uniquely determined by some initial data $\mathcal{Z}_0$. Thus, for simple materials
in the sense of \cite{Noll1972} we obtain the following:
\begin{equation}\label{GenerDefinit}
\mathbf{T} (t) = \mathop{ \mathbf{T} }_{t_0 \leq t' \leq t}  \big(  \mathbf{F}( t'), \mathcal{Z}_0 \big).
\end{equation}
Here, the right-hand side represents a response functional of the material.\footnote{
Note that a \emph{semi-infinite} deformation history was originally
introduced by \cite{Noll1958} without any initial conditions.
Nevertheless, we follow \cite{Noll1972} and consider a \emph{limited} deformation history in \eqref{GenerDefinit},
since it provides the correct framework for the classical plasticity models.}
In certain applications, the data $\mathcal{Z}_0$
may be given by the set of initial values of internal variables.
In general, $\mathcal{Z}_0$ can encapsulate both scalar and tensor-valued quantities.

\begin{figure}\centering
\psfrag{A}[m][][1][0]{$\tilde{\mathcal{K}}$}
\psfrag{B}[m][][1][0]{$\tilde{\mathcal{K}}^{\text{new}}$}
\psfrag{D}[m][][1][0]{$\mathcal{K}$}
\psfrag{E}[m][][1][0]{$\mathbf F_{0}$}
\psfrag{G}[m][][1][0]{$\mathbf F^{\text{new}}$}
\psfrag{K}[m][][1][0]{$\mathbf F$}
\scalebox{0.95}{\includegraphics{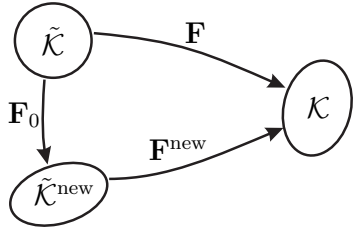}}
\caption{Commutative diagram: change of the reference configuration $\tilde{\mathcal{K}}$
to the new reference $\tilde{\mathcal{K}}^{\text{new}}$.
\label{fig1}}
\end{figure}

Now let us consider a mapping $\textbf{F}_0 = const$, such that $\det(\textbf{F}_0)=1$.
Let $\widetilde{\mathcal{K}}^{\text{new}} := \textbf{F}_0 \widetilde{\mathcal{K}}$
be a new reference configuration (cf. Fig. \ref{fig1}).
We define the new deformation gradient (relative deformation gradient) by
the following push-forward operation
\begin{equation}\label{ShiftNonlinear}
\mathbf{F}^{\text{new}}(t) := \mathbf{F}(t) \ \mathbf{F}_0^{-1}.
\end{equation}
We say that the material model \eqref{GenerDefinit} is weakly invariant under the
transformation \eqref{ShiftNonlinear} when there is a new set of initial data
\begin{equation}\label{NewInitialData}
\mathcal{Z}_0^{\text{new}} = \mathcal{Z}_0^{\text{new}}(\mathcal{Z}_0,\mathbf{F}_0)
\end{equation}
such that the same mechanical response is predicted:
\begin{equation}\label{GenerDefinit2}
\mathop{\mathbf{T}}_{t_0 \leq t' \leq t} \big( \mathbf{F}(t'), \mathcal{Z}_0 \big) =
\mathop{\mathbf{T}}_{t_0 \leq t' \leq t} \big( \mathbf{F}^{\text{new}}(t'), \mathcal{Z}_0^{\text{new}} \big).
\end{equation}
If a certain model is invariant under arbitrary volume-preserving reference changes,
we may be free in our use of the terminology and say that the model is weakly invariant, or even w-invariant.

\textbf{Remark 1.}
Let us recall the classical ``strong'' invariance requirement, which operates with a semi-infinite deformation history
\citep{Noll1958, TruesdNoll}
\begin{equation}\label{GenerDefinitNoll}
\mathop{\mathbf{T}}_{-\infty < t' \leq t} \big( \mathbf{F}(t') \big) =
\mathop{\mathbf{T}}_{-\infty < t' \leq t} \big( \mathbf{F}^{\text{new}}(t') \big).
\end{equation}
The strong invariance \eqref{GenerDefinitNoll} under arbitrary isochoric transformations of
the reference would immediately imply a fluid-like material behavior.
Thus, the strong invariance under arbitrary transformations is too
restrictive for elasto-plasticity.
In contrast to \eqref{GenerDefinitNoll}, the invariance
relations \eqref{GenerDefinit2} are ``weak'' in the sense that
the new initial conditions can counteract the reference change.
Owing to this ``weakness'', a reasonable analysis of elasto-plasticity is possible with this concept.

In order to prevent any misconception, a clear distinction should be made between the
weak invariance \eqref{NewInitialData}, \eqref{GenerDefinit2}
and the strong invariance \eqref{GenerDefinitNoll}.
The difference between both concepts may disappear in case of constitutive relations without
initial conditions.
 $\blacksquare$

\textbf{Remark 2.}
Since the Kirchhoff stress is given by $\mathbf{S} = \det(\mathbf{F}) \mathbf{T}$ and
$ \det(\mathbf{F}) \stackrel{\eqref{ShiftNonlinear}}{=} \det(\mathbf{F}^{\text{new}})$, the same Kirchhoff stress is predicted
upon the reference change by any w-invariant model. $\blacksquare$

\textbf{Remark 3.}
Note that the velocity gradient $\textbf{L}$, the strain rate  $\textbf{D}$
and the continuum spin $\textbf{W}$ are reference independent
(in that sense they represent truly Eulerian quantities)
\begin{equation}\label{InvarianOfStrain}
\textbf{L}^{\text{new}} =
\dot{\mathbf{F}}^{\text{new}} ( \mathbf{F}^{\text{new}} )^{-1} \stackrel{\eqref{ShiftNonlinear}}{=}
\dot{\mathbf{F}} \mathbf{F}^{-1} = \textbf{L},
\end{equation}
\begin{equation}\label{InvarianOfStrain2}
\textbf{D}^{\text{new}} = \text{sym} (\textbf{L}^{\text{new}}) \stackrel{\eqref{InvarianOfStrain}}{=}
\text{sym} (\textbf{L}) = \textbf{D}, \quad
\end{equation}
\begin{equation}\label{InvarianOfStrain3}
\textbf{W}^{\text{new}} = \text{skew} (\textbf{L}^{\text{new}}) \stackrel{\eqref{InvarianOfStrain}}{=}
\text{skew} (\textbf{L}) = \textbf{W}.
\end{equation}
All material models in the form
\begin{equation}\label{GenerDefinitdfsafsds}
\mathbf{T} (t) = \mathop{\mathbf{T}}_{t_0 \leq t' \leq t} \big(  \mathbf{L}( t'), \det(\mathbf{F}(t)), \mathcal{Z}_0 \big)
\end{equation}
can be represented in a weakly invariant form, if
$\mathcal{Z}_0$ is not affected by the reference change
($\mathcal{Z}^{\text{new}}_0 = \mathcal{Z}_0$).
A non-trivial example will be considered in Section 3.2.\footnote{Another example is given by
the new model of finite strain elasto-plasticity, which was
proposed recently by \cite{Volokh2013}.} $\blacksquare$

The notion of w-invariance can be formulated on the reference configuration as well. Indeed, suppose that
 the principle of material frame-indifference holds. Following \cite{TruesdNoll}, the material model \eqref{GenerDefinit}
takes the reduced form with respect to the 2nd Piola-Kirchhoff stress tensor $\widetilde{\mathbf{T}}$
and the right Cauchy-Green
tensor $\mathbf{C}$:
\footnote{Observe that the reduced form \eqref{GenerDefinit3}
corresponds to the \emph{active} interpretation of the principle of frame-indifference.
Thus, indifference with respect to superimposed rigid-body motions is postulated
as a constitutive assumption \citep{Svendsen1999, BertramSvendsen2001}.}
\begin{equation}\label{GenerDefinit3}
\widetilde{\mathbf{T}} (t) =
\mathop{\widetilde{\mathbf{T}}}_{t_0 \leq t' \leq t} \big( \mathbf{C}(t'), \mathcal{Z}_0 \big), \
\text{where} \ \mathbf{C}(t') := \mathbf{F}^{\text{T}}(t') \mathbf{F}(t').
\end{equation}
The model \eqref{GenerDefinit3} is weakly invariant under the reference change if
there is a new set $\mathcal{Z}_0^{\text{new}} = \mathcal{Z}_0^{\text{new}}(\mathcal{Z}_0,\mathbf{F}_0)$ such that
\begin{equation}\label{GenerDefinit4}
\mathop{\widetilde{\mathbf{T}}}_{t_0 \leq t' \leq t} \big(\mathbf{C}(t'), \mathcal{Z}_0 \big)  =
\mathbf{F}_0^{-1} \mathop{\widetilde{\mathbf{T}}}_{t_0 \leq t' \leq t} \big(
\mathbf{C}^{\text{new}}(t'), \mathcal{Z}_0^{\text{new}} \big)
\mathbf{F}_0^{-\text{T}},
\end{equation}
where the new right Cauchy-Green tensor is obtained using the push forward operation
\begin{equation}\label{TransformatRightCauch}
\mathbf{C}^{\text{new}} (t) := \mathbf{F}_0^{-\text{T}} \mathbf{C} (t) \mathbf{F}_0^{-1}.
\end{equation}
The 2nd Piola-Kirchhoff stress must obey
\eqref{GenerDefinit4} in order to ensure that
the true stresses remain invariant.
An example of such model will be given in Section 5.

Dealing with constitutive equations enjoying the w-invariance, the reference
configuration can be chosen in a rather arbitrary manner. The only restriction
is imposed by the incompressibility condition $\det(\textbf{F}_0)=1$.

Two physical observations are helpful for a better understanding of the w-invariance:
\begin{itemize}
\item[(p)] Whereas the strong invariance under arbitrary isochoric reference changes is observed only for fluids, the plastic flow of solids
may also exhibit some fluid-like features, which are formalized here using w-invariance.
\item[(e)] The current deformation of the crystal lattice and not the overall deformation of the metallic
material determine the Cauchy stresses \citep{Anand87, Bertram1998, Rubin2001}.
\end{itemize}
Observations (p) and (e) impose some restrictions on the plastic and elastic constitutive relations, respectively.
These observations are complimentary to each other and they motivate the
current study from a physical standpoint.


\section{Additive hypoelasto-plasticity}

\subsection{General framework of the additive hypoelasto-plasticity}

As a first example, let us consider the classical approach to elasto-plasticity (cf. \cite{Neale1981, Nemat-Nasser1982}) which is based on the additive decomposition
of the strain rate tensor $\mathbf{D}$ into the elastic part $\mathbf{D}_{\text{e}}$
and the plastic part $\mathbf{D}_{\text{p}}$:
\begin{equation}\label{AdditivDecomp}
\mathbf{D} = \mathbf{D}_{\text{e}} + \mathbf{D}_{\text{p}}.
\end{equation}
The elastic part $\mathbf{D}_{\text{e}}$ is connected to a certain
objective rate of the Kirchhoff stress tensor via hypoelastic relations
\begin{equation}\label{Hypo}
\mathbf{D}_{\text{e}} = \mathbb{H} : \accentset{o}{\mathbf S},
\end{equation}
where $\mathbb{H}$ is a fourth rank compliance tensor, $\mathbf{S}$ is the Kirchhoff
stress.\footnote{Equation \eqref{Hypo} corresponds to the so-called grade-zero hypoelasticity.
It is a special case of relations of Eulerian rate type (cf. \cite{Eshraghi2013a, Eshraghi2013b}).}
For simplicity,
we suppose here that $\mathbb{H}$ is constant.
Next, the Mises-Huber yield function is considered
\begin{equation}\label{MisHuberYield}
f := \| \mathbf{S}^{\text{D}} \| - \sqrt{2/3} K,
\end{equation}
where $K > 0$ stands for the yield stress.\footnote{Note
that the yield function is formulated in terms of the Kirchhoff stress, which
is believed to be a natural choice for metals.
For other groups of materials, like certain geomaterials, the true stresses
can be used (cf. \cite{Yamakawa2010}).}
The flow rule is given by
\begin{equation}\label{FLOWRULE}
\mathbf{D}_{\text{p}} = \lambda_{\text{p}}
\frac{\displaystyle \mathbf{S}^{\text{D}}}{\displaystyle \| \mathbf{S}^{\text{D}} \| }.
\end{equation}
Here, $ \lambda_{\text{p}} \geq 0$ stands for the plastic multiplier. The Kuhn-Tucker conditions
take the same form as in the small strain case
\begin{equation}\label{Hypoddfs}
f \leq 0, \ \lambda_{\text{p}} \geq 0, \ f \lambda_{\text{p}} =0.
\end{equation}
The initial conditions are imposed on the Kirchhoff stresses
\begin{equation}\label{IntCondKirch}
\mathbf{S}|_{t=t_0} = \mathbf{S}_0, \quad
\mathcal{Z}_0 := \{ \mathbf{S}_0 \}.
\end{equation}
Note that the objective stress rate which appears in \eqref{Hypo} has to be specified.
In this work we imply the consistency criterion proposed by \cite{Prager1960}
in order to restrict the class of possible stress rates.
The consistency criterion states that
$\accentset{o}{\mathbf S} = \mathbf{0}$ implies $f = const$, where the yield function $f$
is given by \eqref{MisHuberYield}.
In order to satisfy this condition,
the stress rate $\accentset{o}{\mathbf S}$ must be a corotational derivative
(cf. \cite{XiaoBruhnsMeyers2000, BruhnsMeyersXiao2004})
\begin{equation}\label{GenerCorotDerivat}
\accentset{o}{\mathbf S} := \dot{ \mathbf{S}} +  \mathbf{S} \mathbf{\Omega} - \mathbf{\Omega}  \mathbf{S}, \quad
\mathbf{\Omega} \in \text{Skew}.
\end{equation}
Here, Skew stands for the set of skew-symmetric tensors,
the skew-symmetric operator $\mathbf{\Omega}$ is referred to as a spin tensor, superimposed dot
stands for the material time derivative.
There are infinitely many ways of defining the spin tensors $\mathbf{\Omega}$
\citep{Xiao1998, Korobeynikov, HashiguchiBook}.
Obviously, the properties of the resulting constitutive
equations are strongly dependent on the specific choice of $\mathbf{\Omega}$.
In particular, we have the following theorem.

\textbf{Theorem 1.}
\emph{Constitutive relations \eqref{AdditivDecomp}---\eqref{GenerCorotDerivat}
are weakly invariant under the isochoric reference change if and only if
the spin tensor $\mathbf{\Omega}$ remains invariant under such a change.}

\textbf{Proof:}
After some algebraic computations, the constitutive equations can be
rewritten in the compact form (cf. equation (45) in \cite{BruhnsXiao1999}):
\begin{equation}\label{CompactForm}
\accentset{o}{\mathbf S} = \mathfrak{F}(\mathbf{S}, \mathbf{D}),
\end{equation}
where the right-hand side depends solely on $\mathbf{S}$ and $\mathbf{D}$.\footnote{
Equation \eqref{CompactForm} is a remarkable result since
the plastic part $\mathbf{D}_{\text{p}}$ is exluded.}
Consider a local loading process $\mathbf{F}( t' ), t' \in [t_0, t]$ and let
$\textbf{F}_0 = const$ such that $\det(\textbf{F}_0)=1$.
As already mentioned in Section 2, the material model \eqref{CompactForm} is w-invariant if and only if
it predicts the same Kirchhoff stress $\mathbf{S}$ for
the new process $ \mathbf{F}( t' ) \mathbf{F}_0^{-1}, t' \in [t_0, t]$.
According to \eqref{IntCondKirch}, the set of initial data is
given by $\mathcal{Z}_0 = \{ \mathbf{S}_0 \}$.
Thus, the transformation rule \eqref{NewInitialData} must take the following trivial form
\begin{equation}\label{TrivialTransformation}
\mathcal{Z}^{\text{new}}_0 = \mathcal{Z}_0 = \{ \mathbf{S}_0 \}
\end{equation}
which is independent of $\textbf{F}_0$.

Assume that the model \eqref{CompactForm} is w-invariant. We need to show
that the spin tensor $\mathbf{\Omega}$
is invariant (reference independent).
First, recall that the strain rate tensor $\mathbf{D}$ fulfils the invariance.
Thus, the right-hand side of \eqref{CompactForm} represents an invariant quantity.
Therefore, \eqref{CompactForm} implies that $\accentset{o}{\mathbf S}$ is invariant as well.
Taking $\eqref{GenerCorotDerivat}_1$ into account, we conclude that
\begin{equation}\label{InvarianOfStrain33}
\mathbf{S} \mathbf{\Omega}^{\text{new}} - \mathbf{\Omega}^{\text{new}}  \mathbf{S} = \mathbf{S} \mathbf{\Omega} - \mathbf{\Omega}  \mathbf{S}.
\end{equation}
Here, the spin tensors $\mathbf{\Omega}$ and $\mathbf{\Omega}^{\text{new}}$ correspond to loading
histories $\mathbf{F}( t' )$ and
$\mathbf{F}( t' ) \mathbf{F}_0^{-1}$ respectively.
Observe that arbitrary initial conditions \eqref{IntCondKirch}
can be imposed on the Kirchhoff stresses.\footnote{The only restriction is imposed by $\| \mathbf{S}_0^{\text{D}} \| \leq \sqrt{2/3} K$.}
Therefore, \eqref{InvarianOfStrain33} must hold for all symmetric $\mathbf{S}$.
For the difference $\mathbf{\Omega}^{\text{dif}}:= \mathbf{\Omega}^{\text{new}} - \mathbf{\Omega} \in \text{Skew}$ we obtain
from \eqref{InvarianOfStrain33}
\begin{equation}\label{InvarianOfStrain4}
\mathbf{S} \ \mathbf{\Omega}^{\text{dif}} = \mathbf{\Omega}^{\text{dif}} \  \mathbf{S} \ \ \text{for all} \ \ \mathbf{S} \in \text{Sym}.
\end{equation}
Thus, for all vectors $\mathbf{x}$ we have
\begin{equation}\label{InvarianOfStrain5}
(\mathbf{x} \otimes \mathbf{x}) \ \mathbf{\Omega}^{\text{dif}} = \mathbf{\Omega}^{\text{dif}} \  (\mathbf{x} \otimes \mathbf{x}).
\end{equation}
Therefore, obviously,
\begin{equation}\label{InvarianOfStrain6}
\mathbf{x} \otimes ( \mathbf{x}  \mathbf{\Omega}^{\text{dif}} ) = ( \mathbf{\Omega}^{\text{dif}}  \mathbf{x} ) \otimes \mathbf{x}.
\end{equation}
But, on the other hand, $\mathbf{\Omega}^{\text{dif}} \in \text{Skew}$ yields
\begin{equation}\label{InvarianOfStrain7}
\mathbf{x}  \mathbf{\Omega}^{\text{dif}} = (\mathbf{\Omega}^{\text{dif}})^{\text{T}} \ \mathbf{x} = -\mathbf{\Omega}^{\text{dif}} \ \mathbf{x}.
\end{equation}
Substituting this into \eqref{InvarianOfStrain6}, we arrive at the following for all $\mathbf{x}$
\begin{equation}\label{InvarianOfStrain8}
 - \mathbf{x} \otimes (\mathbf{\Omega}^{\text{dif}}  \mathbf{x}) = ( \mathbf{\Omega}^{\text{dif}}  \mathbf{x} ) \otimes \mathbf{x}.
\end{equation}
Thus, $\alpha_{\mathbf{x}} \in \mathbb{R}$ exists such that $\mathbf{\Omega}^{\text{dif}}  \mathbf{x} = \alpha_{\mathbf{x}}  \mathbf{x}$ and
from \eqref{InvarianOfStrain8} we come to
\begin{equation}\label{InvarianOfStrain9}
 - \alpha_{\mathbf{x}} \ \mathbf{x} \otimes \mathbf{x} = \alpha_{\mathbf{x}} \ \mathbf{x} \otimes \mathbf{x}.
\end{equation}
For all $\mathbf{x} \neq 0$, we have $\alpha_{\mathbf{x}}=0$ and
\begin{equation}\label{InvarianOfStrain10}
\mathbf{\Omega}^{\text{dif}}  \mathbf{x} = \mathbf{0}, \quad
\mathbf{\Omega}^{\text{dif}}  = \mathbf{0}.
\end{equation}
In other words,  relation \eqref{InvarianOfStrain33} implies $\mathbf{\Omega}^{\text{dif}} = \mathbf{\Omega}^{\text{new}} - \mathbf{\Omega} = \mathbf{0}$.
Hence, \eqref{InvarianOfStrain33} holds only if
$\mathbf{\Omega}^{\text{new}} = \mathbf{\Omega}$.

For the proof in the opposite direction, assume
that the spin $\mathbf{\Omega}$ fulfils the invariance.
In which case, obviously, the model \eqref{CompactForm} is w-invariant.
The theorem is proved $\blacksquare$

Different models of elasto-plasticity can be constructed by an appropriate choice of the spin tensor \
$\mathbf{\Omega}$.\footnote{The reader interested in the numerical treatment of
constitutive equations for a variety of different spin tensors is referred to \cite{AbbasiParsa2006}.}
Let us consider some of the most popular options.

\subsection{Zaremba-Jaumann and Green-Naghdi rates}

Probably, the most simple choice for the spin tensor $\mathbf{\Omega}$ is the continuum
spin $\mathbf{W}$, which yields the Zaremba-Jaumann rate, also
known as Zaremba-Jaumann-Noll rate
\begin{equation}\label{ZJRATE}
\mathbf{\Omega}^{ZJ} := \mathbf{W} = \text{skew}(\mathbf{L}), \quad
\accentset{o}{\mathbf S}^{ZJ} := \dot{ \mathbf{S}} +  \mathbf{S} \mathbf{\Omega}^{ZJ} - \mathbf{\Omega}^{ZJ}  \mathbf{S}.
\end{equation}
Since the continuum spin $\mathbf{W}$ is
invariant (recall Remark 3), \emph{the corresponding system of equations is w-invariant} as well.
One important drawback of such an approach is that the stress response exhibits
non-physical oscillations under the simple shear (\cite{Lehmann1972, Dienes1979}).
In order to prevent the oscillations of stresses, the so-called Green-Naghdi
rate\footnote{It is also known as Green-Naghdi-Dienes rate, Green-McInnis rate or polar rate.}
can be implemented. Its definition makes use of the rotation tensor $\mathbf{R}$ which
results from the polar decomposition of the deformation gradient:
\begin{equation}\label{GNRATE}
\mathbf{\Omega}^{GN} := \dot{\mathbf{R}} \mathbf{R}^{\text{T}} \in \text{Skew}, \quad
\accentset{o}{\mathbf S}^{GN} := \dot{ \mathbf{S}} +  \mathbf{S} \mathbf{\Omega}^{GN} - \mathbf{\Omega}^{GN}  \mathbf{S}.
\end{equation}
It can be easily shown that this spin tensor depends on the choice of the
reference configuration (cf. \cite{Korobeynikov2008} ).
Indeed, in general we have
\begin{equation}\label{nonequal}
\mathbf{\Omega}^{GN} \neq \mathbf{W}.
\end{equation}
Therefore, a loading history and a time instance $t'$ exist such that
$\mathbf{\Omega}^{GN}(t') \neq \mathbf{W}(t')$.
By choosing $\mathbf{F}_0 := \overline{\mathbf{F}(t')}$, where
$\overline{\mathbf{X}} := (\det \mathbf{X} )^{-1/3} \mathbf{X}$, we obtain
\begin{equation}\label{newDefGradFor}
\mathbf{F}^{\text{new}} (t') = \mathbf{F} (t') (\overline{\mathbf{F} (t')})^{-1} =
(\det\mathbf{F}(t'))^{1/3}  \mathbf{1}.
\end{equation}
In this special case, the new GN-spin coincides with the continuum spin:
\begin{equation}\label{wildInequal}
(\mathbf{\Omega}^{GN})^{\text{new}}(t') = \mathbf{W}(t') \neq \mathbf{\Omega}^{GN}(t').
\end{equation}
Thus, in general,
\begin{equation}\label{GNRATE2}
(\mathbf{\Omega}^{GN})^{\text{new}} \neq \mathbf{\Omega}^{GN}.
\end{equation}
In other words, $\mathbf{\Omega}^{GN} = \dot{\mathbf{R}} \mathbf{R}^{\text{T}}$
is not a purely Eulerian quantity, since it depends on the choice of the reference configuration.
Hence, according to Theorem 1, the \emph{constitutive equations based on \eqref{GNRATE} are not w-invariant}.

\textbf{Remark 4.} The spin $\mathbf{\Omega}^{GN}$ fulfills invariance under the rigid
rotation of the reference configuration.
Indeed, for $\mathbf{F}_0 = \mathbf{Q}_0 \in \text{Orth}$, we get
$\mathbf{R}^{\text{new}}= \mathbf{R} \mathbf{Q}_0^{-1}$. Therefore,
$\dot{\mathbf{R}}^{\text{new}} (\mathbf{R}^{\text{new}})^{-1} =
\dot{\mathbf{R}}  \mathbf{Q}_0^{-1} \mathbf{Q}_0 \mathbf{R}^{-1} = \dot{\mathbf{R}} \mathbf{R}^{-1}$
for $\mathbf{F}_0 = \mathbf{Q}_0 \in \text{Orth}$.
Thus, initially isotropic materials can be modeled using this stress rate. $\blacksquare$

\begin{figure}\centering
\psfrag{A}[m][][1][0]{$ \mathbf{T}_{1 3}$}
\psfrag{B}[m][][1][0]{$ \mathbf{T}_{2 3}$}
\psfrag{T}[m][][1][0]{time $t$}
\scalebox{0.7}{\includegraphics{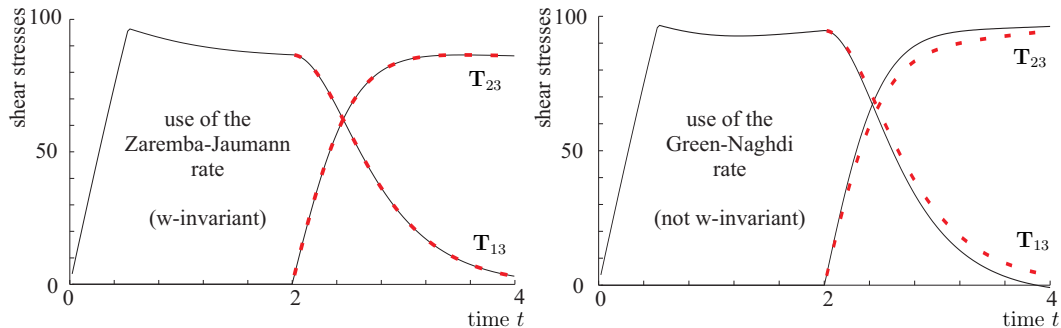}}
\caption{Simulated shear stresses under double simple shear. Unified simulations for $t \in [0,4]$ are depicted by solid lines.
Results obtained after a restart at $t=2$ with a new reference configuration
are shown by dotted lines.
Models which are weakly invariant under the reference change predict the same results
for the restarted simulation (left). Otherwise, the stress response is affected by
the choice of the reference configuration (right).  \label{fig2}}
\end{figure}

Now let us illustrate the implications of w-invariance using a simple numerical example.
To be definite, the compliance tensor $\mathbb{H}$ which appears in \eqref{Hypo} is defined here through its inverse
\begin{equation}\label{HypoConcreteExp}
\mathbb{H}^{-1} : \mathbf{X} = \frac{\displaystyle E}{\displaystyle 3 (1 - 2 \nu)} \text{tr}(\mathbf{X}) \ \mathbf{1}
+  \frac{\displaystyle E}{1 + \nu} \mathbf{X}^{\text{D}}, \ \text{for all} \ \mathbf{X} \in \text{Sym}.
\end{equation}
We insert the values $E = 500$ and $\nu = 0.3$ (all quantities are non-dimensional). The yield stress is given by $K= \sqrt{3} \cdot 100$.
First, we simulate a simple shear with an abrupt load path change according
to the following loading program:
\begin{equation}\label{LoadProgram}
\mathbf{F}(t) =  \begin{cases}
    \mathbf{1}  + t \ \mathbf{e}_1 \otimes \mathbf{e}_3  \quad  \quad \quad \quad  \quad \quad \quad  \quad \quad \ \ \ \  \ \,  \text{if} \ t \in [0,2) \\
    \big(\mathbf{1}  + (t-2) \ \mathbf{e}_2 \otimes \mathbf{e}_3 \big) \big(\mathbf{1}  + 2 \mathbf{e}_1 \otimes \mathbf{e}_3 \big) \quad \text{if} \ t \in [2,4]
\end{cases},
\end{equation}
where $\{ \mathbf{e}_1, \mathbf{e}_2, \mathbf{e}_3 \}$ is a fixed orthonormal basis. Since $\mathbf{F}(0) =  \mathbf{1}$, the initial configuration coincides
with the reference configuration here.
As an initial condition at $t=0$ we assume zero stresses.
The simulated stress response is shown by solid lines in Fig. \ref{fig2} (left) for the Zaremba-Jaumann model and
in Fig. \ref{fig2} (right) for the Green-Naghdi model.

Now let us restart the simulation at the time instance $t=2$ using a new
reference configuration $\widetilde{\mathcal{K}}^{\text{new}} = \textbf{F}_0 \widetilde{\mathcal{K}}$
where $\mathbf{F}_0 = \textbf{F}(2) = \mathbf{1}  + 2 \mathbf{e}_1 \otimes \mathbf{e}_3$.
In other words, the new reference configuration corresponds to the configuration occupied by the body at $t=2$.
The corresponding initial conditions (at $t=t_0=2$) are formulated with respect to the Kirchhoff stresses.
The initial values are taken from the simulation results obtained in the previous simulation.
Within the time interval $t \in [2,4] $, the material is subjected to the simple shear
\begin{equation}\label{NewDefGradien}
\mathbf{F}^{\text{new}}(t) \stackrel{\eqref{ShiftNonlinear}}{=}  \mathbf{F}(t) \textbf{F}_0^{-1}
\stackrel{\eqref{LoadProgram}}{=} \mathbf{1}  + (t-2) \ \mathbf{e}_2 \otimes \mathbf{e}_3.
\end{equation}
The simulation results are shown in Figure \ref{fig2} by dotted lines.
Observe that exactly the same stress response is predicted by the Zaremba-Jaumann model after restart, since this model is
w-invariant under arbitrary isochoric changes of the reference configuration. For the Green-Naghdi model, in contrast,
a divergent response is obtained.

The set of initial data at $t=t_0=2$ is given by the initial value
of the Kirchhoff stress tensor, but \emph{these data are not sufficient to describe
the material state in a unique manner} if the model is not w-invaraint.
For such models, additional information
concerning the concrete choice of the reference is required.

\subsection{Models based on the logarithmic rate}

A common drawback of the models relying on the Zaremba-Jaumann and Green-Naghdi rates is
that the constitutive equations are not integrable within the elastic range.
In other words, even for $\mathbf{D} = \mathbf{D}_{\text{e}}$
the stress state depends on the loading path, which
is inconsistent with the classical notion of elasticity \citep{SimoPister1984}.
Moreover, it was proved by \cite{BruhnsXiao1999} that, among all possible models of
the corotational rate type with constant elastic moduli, the
model based on the logarithmic rate \emph{is the only choice} which allows for
obtaining integrable stress-strain relations
and avoiding the non-physical energy-dissipation
in the elastic range. For that reason, the logarithmic
rate deserves serious consideration.

The corotational logarithmic rate
is given by
\begin{equation}\label{Equation}
\accentset{o}{\mathbf S}^{\text{log}} := \dot{ \mathbf{S}} +  \mathbf{S} \mathbf{\Omega}^{\text{log}} -
\mathbf{\Omega}^{\text{log}}  \mathbf{S},
\end{equation}
where $\mathbf{\Omega}^{\text{log}}$ stands for
the so-called log-spin \citep{XiaoBruhns1997}. This logarithmic rate has the following
remarkable property: The strain rate $\mathbf{D}$ equals the logarithmic rate of the Eulerian logarithmic strain
measure $\ln \mathbf{V}$
\begin{equation}\label{RemarkProperty}
\mathbf{D} =    \accentset{o}{(\ln \mathbf{V})}^{\text{log}}, \quad \mathbf{V} := \sqrt{\mathbf{F} \mathbf{F}^{\text{T}}  }.
\end{equation}

The logarithmic rate is uniquely determined by the kinematic requirement $\eqref{RemarkProperty}_1$.
Property $\eqref{RemarkProperty}_1$ has become the basis for several advanced models of plasticity, including models with
plastic anisotropy and models for shape memory alloys (cf. \cite{MuellerBruhns2006, Arghavani2011,
Naghdabadi2012, Teeriaho2013, Xiao2013, Zhu2014}).
For the discussion of this seminal idea in historical perspective the reader is referred to \cite{XiaoReviw} and \cite{Zhilin}.
The closed form coordinate-free expression for the logarithmic spin tensor $\mathbf{\Omega}^{\text{log}}$
is given in Appendix A. The log-spin $\mathbf{\Omega}^{\text{log}}$ is a smooth
function of the strain rate $\mathbf{D}$, the continuum spin
$\mathbf{W}$ and the left Cauchy-Green tensor $\mathbf{B}:= \mathbf{F} \mathbf{F}^{\text{T}}$
\begin{equation}\label{LogarithmSpin}
\mathbf{\Omega}^{\text{log}} = \mathbf{\Omega}^{\text{log}} (\mathbf{D}, \mathbf{W},\mathbf{B}) \in \text{Skew}.
\end{equation}

All arguments are essential in \eqref{LogarithmSpin}. Observe that the first two arguments
are invariant under the reference change,
but the left Cauchy-Green tensor $\mathbf{B}$ is not.
Let us show that the \emph{log-spin tensor depends on the choice of the reference configuration}.
In general, $\mathbf{\Omega}^{\text{log}} (\mathbf{D}, \mathbf{W},\mathbf{B}) \neq \mathbf{W}$, but
in the special case where $\mathbf{B}$ is isotropic, the log-spin coincides with the continuum spin (cf. Appendix A):
\begin{equation}\label{LogarithmSpin2}
\mathbf{\Omega}^{\text{log}} (\mathbf{D}, \mathbf{W},\mathbf{B}) = \mathbf{W}, \ \text{if} \ \mathbf{B} = \beta \mathbf{1}, \beta >0.
\end{equation}
Let us consider a time instance $t'$ and a corresponding set $\{ \mathbf{D}, \mathbf{W},\mathbf{B} \}$ such
that $\mathbf{\Omega}^{\text{log}} (\mathbf{D}, \mathbf{W},\mathbf{B}) \neq \mathbf{W}$.
By choosing $\mathbf{F}_0 := \overline{\mathbf{F}(t')}$, we arrive
at $t=t'$
\begin{equation}\label{LogarithmSpin3}
\mathbf{F}^{\text{new}} (t') = \mathbf{F} (t') (\overline{\mathbf{F}
(t')})^{-1} = (\det\mathbf{F}(t'))^{1/3}  \mathbf{1}, \quad
\mathbf{B}^{\text{new}}(t') = (\det\mathbf{F}(t'))^{2/3}  \mathbf{1}.
\end{equation}
Thus, at $t=t'$, we have
\begin{equation}\label{LogarithmSpin4}
( \mathbf{\Omega}^{\text{log}} )^{\text{new}} (t') \stackrel{\eqref{LogarithmSpin2}}{=} \mathbf{W}(t') \neq \mathbf{\Omega}^{\text{log}}(t').
\end{equation}
Therefore, the log-spin $\mathbf{\Omega}^{\text{log}}$ depends on the choice of the reference configuration.\footnote{
Observe that $\ln \mathbf{V}$ and $\mathbf{\Omega}^{\text{log}}$ are reference dependent. As it was pointed out by an anonymous reviewer,
the reference dependencies counteract each other so that $\accentset{o}{(\ln \mathbf{V})}^{\text{log}}$
becomes reference independent, which is a noticeable property of the log-spin.}
It follows from Theorem 1 that the \emph{model based on the logarithmic rate is not w-invariant}.


\textbf{Remark 5.} The logarithmic spin meets invariance with respect to rigid
rotation of the reference configuration.
Indeed, for $\mathbf{F}_0 = \mathbf{Q}_0 \in \text{Orth}$, we get
$\mathbf{B}^{\text{new}}= \mathbf{B}$. Therefore,
$(\mathbf{\Omega}^{\text{log}})^{\text{new}}  \stackrel{\eqref{LogarithmSpin}}{=}
\mathbf{\Omega}^{\text{log}}$ for $\mathbf{F}_0 = \mathbf{Q}_0 \in \text{Orth}$.
Hence, initially isotropic materials can be modeled using this stress rate.
Moreover, constitutive modeling of initially
anisotropic materials is also possible, as discussed in \cite{Xiao2007}  $\blacksquare$

\section{Additive elasto-plasticity in the logarithmic strain space}

Now we proceed to another popular class of constitutive equations. Following \cite{Papadopoulus, Miehe2002, Schroeder2002},
we consider a framework which adopts the structure
of the the geometrically linear relations \eqref{Prandtl-Reuss} -- \eqref{IniCond}.
In these relations, the linearized strain tensor $\boldsymbol{\varepsilon}(t)$
is formally replaced by the Lagrangian logarithmic strain (Hencky strain)
\begin{equation}\label{LagrLog}
\mathbf{H}(t) := \frac{1}{2} \ln (\mathbf{C}(t)).
\end{equation}
The resulting stress response $\boldsymbol{\sigma}$ is now understood as
a certain Lagrangian stress measure which is power conjugate to the logarithmic
strain $\mathbf{H}$
\begin{equation}\label{PowerConjugate}
\boldsymbol{\sigma} : \dot{\mathbf{H}} = \widetilde{\textbf{T}} : \Big(\frac{1}{2} \dot{\mathbf{C}} \Big) \ \text{for all} \ \dot{\mathbf{C}} \in Sym.
\end{equation}
Thus, $\boldsymbol{\sigma}$ and $\mathbf{H}$ are dual variables.
In order to obtain the relation between $\boldsymbol{\sigma}$ and the conventional stress
measures we rewrite \eqref{PowerConjugate} in the form
\begin{equation}\label{PowerConjugate2}
\boldsymbol{\sigma} : \Big(\frac{1}{2} \frac{\partial \ln (\mathbf{C})}{\partial \mathbf{C}} : \dot{\mathbf{C}} \Big)
= \widetilde{\textbf{T}} : \Big(\frac{1}{2} \dot{\mathbf{C}} \Big) \ \text{for all} \ \dot{\mathbf{C}} \in Sym.
\end{equation}
Next, taking into account that the derivative $ \frac{\partial \ln (\mathbf{C})}{\partial \mathbf{C}}$
is super-symmetric (cf. \cite{ItskovBook}), we obtain\footnote{Numerical procedures for the computation of
conventional stresses basing on $\boldsymbol{\sigma}$ were discussed,
among others, by \cite{Plesek2006} and \cite{ItskovBook}.}
\begin{equation}\label{PowerConjugate3}
\Big( \frac{\partial \ln (\mathbf{C})}{\partial \mathbf{C}} : \boldsymbol{\sigma} \Big) : \dot{\mathbf{C}}
= \widetilde{\textbf{T}} : \dot{\mathbf{C}} \ \text{for all} \ \dot{\mathbf{C}} \in Sym.
\end{equation}
Therefore,
\begin{equation}\label{PowerConjugate4}
\widetilde{\textbf{T}} = \frac{\partial \ln (\mathbf{C})}{\partial \mathbf{C}} : \boldsymbol{\sigma}, \quad
\textbf{T} = \frac{1 }{\det \mathbf{F}} \mathbf{F} \Big(\frac{\partial \ln
(\mathbf{C})}{\partial \mathbf{C}} : \boldsymbol{\sigma} \Big) \mathbf{F}^{\text{T}}.
\end{equation}

Following the terminology of \cite{Miehe2002}, the system of constitutive equations is subdivided into three modules:
\begin{itemize}
\item  equation \eqref{LagrLog} as a geometric preprocessor ($\mathbf{C} \mapsto \mathbf{H}$)
\item  equations \eqref{Prandtl-Reuss} -- \eqref{IniCond} as a constitutive model ($\mathbf{H} \equiv \boldsymbol{\varepsilon} \mapsto \boldsymbol{\sigma}$)
\item  equation $\eqref{PowerConjugate4}_1$ as a geometric postprocessor ($\boldsymbol{\sigma} \mapsto \widetilde{\textbf{T}}$)
\end{itemize}

As an alternative to \eqref{LagrLog}, a series of other generalized strain measures can be
implemented. Here we concentrate on logarithmic strain for the following reason. As deduced by \cite{Itskov2004},
the logarithmic strain yields the most appropriate results in case of rigid-plastic material subjected to simple shear.

\begin{figure}\centering
\psfrag{A}[m][][1][0]{$\| \mathbf{H} (t) \|$}
\psfrag{B}[m][][1][0]{$\|  \mathbf{H}^{\text{new}} (t) - \mathbf{H}^{\text{new}} (0) \|$}
\psfrag{T}[m][][1][0]{$t$}
\scalebox{0.9}{\includegraphics{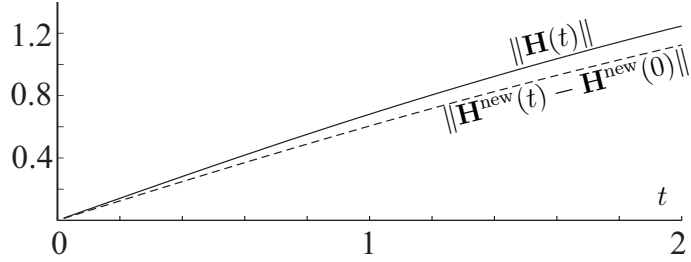}}
\caption{Evolution of the norm of the Hencky strain under
the simple shear \eqref{LoadProgramLog} depending
on the choice of reference configuration.  \label{fig3}}
\end{figure}

Let us show that the corresponding constitutive relations \emph{are not w-invariant under the reference change}.
Toward that end, consider an example of simple shear given by
\begin{equation}\label{LoadProgramLog}
\mathbf{F}(t) = \mathbf{1}  + t \ \mathbf{e}_2 \otimes \mathbf{e}_3.
\end{equation}
For the model under consideration, the initial state is described by
$\mathcal{Z}_0 = \{ \boldsymbol{\varepsilon}^{0}_{\text{p}} \}$ (recall the initial condition \eqref{IniCond}).
To be definite, we assume zero stress at $t=0$
\begin{equation}\label{IntCondLog}
\mathbf{T}|_{t=0} = \mathbf{0}.
\end{equation}
Due to \eqref{Prandtl-Reuss}, \eqref{IsotHooke}, and \eqref{PowerConjugate4}, the initial condition \eqref{IntCondLog} implies
\begin{equation}\label{IntCondLog2}
\boldsymbol{\sigma}|_{t=0} = \mathbf{0}, \quad \boldsymbol{\varepsilon}_{\text{p}}|_{t=0} = \mathbf{H}|_{t=0} = \mathbf{0}.
\end{equation}
Next, consider a reference change which is described by
$\mathbf{F}_0 = \mathbf{1}  +  2 \ \mathbf{e}_1 \otimes \mathbf{e}_3$.
The new load path is then given by
\begin{equation}\label{NewLoadPath}
\mathbf{F}^{\text{new}} (t) = \mathbf{F} (t) \mathbf{F}_0^{-1}, \quad
\mathbf{C}^{\text{new}} (t) = \mathbf{F}_0^{-\text{T}} \mathbf{C} (t) \mathbf{F}_0^{-1}, \quad
\mathbf{H}^{\text{new}}(t) = \frac{1}{2} \ln (\mathbf{C}^{\text{new}}(t)).
\end{equation}
In order to obtain zero stress at $t=0$, the following initial condition
must be imposed on $\boldsymbol{\varepsilon}^{\text{new}}_{\text{p}}$
\begin{equation}\label{IntCondLog3}
\boldsymbol{\varepsilon}^{\text{new}}_{\text{p}}|_{t=0} = \mathbf{H}^{\text{new}}|_{t=0}.
\end{equation}
Let the yield functions $f$ and $f^{\text{new}}$ correspond to processes \eqref{LoadProgramLog}, $\eqref{IntCondLog2}_2$ and
$\eqref{NewLoadPath}_1$, \eqref{IntCondLog3} respectively.
Synchronized yielding is a necessary
condition of w-invariance, which is possible only if
$f(t) = f^{\text{new}}(t)$.
Thus, the following condition is necessary for the w-invariance
\begin{equation}\label{InvarOfNorms}
 \| \boldsymbol{\sigma}^{\text{D}} \| =   \| ( \boldsymbol{\sigma}^{\text{new}})^{\text{D}} \|.
\end{equation}
The lack of the w-invariance becomes evident upon examination of the elasticity relations.
Indeed, assuming that for small $t$ the material remains in the
elastic range, we have
\begin{equation}\label{ElasticRange}
\boldsymbol{\varepsilon}_{\text{p}}(t) = \mathbf{H}|_{t=0} = \mathbf{0}, \quad
\boldsymbol{\varepsilon}^{\text{new}}_{\text{p}}(t) = \mathbf{H}^{\text{new}}|_{t=0}.
\end{equation}
Combining this with \eqref{IsotHooke}, and taking into account that $\text{tr}\mathbf{H} = \text{tr}\mathbf{H}^{\text{new}} = 0$,
the condition \eqref{InvarOfNorms} is reduced to
\begin{equation}\label{NecessConditions}
\| \mathbf{H} (t) \| =   \|  \mathbf{H}^{\text{new}} (t) - \mathbf{H}^{\text{new}} (0) \|.
\end{equation}
As shown in Fig. \ref{fig3}, this necessary condition is violated. Since \eqref{InvarOfNorms} does not hold true,
the time instance of plastification depends on the choice
of the reference configuration. Thus, the constitutive equations
are not w-invariant under the reference change.
Roughly speaking, the transformation of the reference cannot be neutralized by any
transformation of the initial conditions.

\section{Multiplicative hyperelasto-plasticity}

Finally, let us analyse a model of hyperelastic-perfectly-plastic
response, which is covered as a special case by the classical model of \cite{Simo92}. The model is based on the multiplicative decomposition
of the deformation gradient into the elastic part $\hat{\mathbf F}_{\text{e}}$ and the
plastic part $\mathbf F_{\text{p}}$
\begin{equation}\label{Simo1}
\mathbf F = \hat{\mathbf F}_{\text{e}} \ \mathbf F_{\text{p}}.
\end{equation}
This decomposition was originally considered by \cite{Bilby1957} and
\cite{Kroener1958} for metallic materials.\footnote{
As discussed in \cite{Clayton2014, Le2014},
two different mechanical interpretations
of terms ``elastic" and ``plastic"
were suggested by Bilby and Kr\"oner.}
It can be postulated as a phenomenological assumption or it
can be derived from the concept of material isomorphism under some general assumptions \citep{Bertram1998, Bertram2003}.
The decomposition \eqref{Simo1}  gives rise to the weighted stress tensor operating on the
intermediate configuration (also known as stress-free configuration)
\begin{equation}\label{weighinter}
\hat{\mathbf S} : = {\hat{\mathbf F}_{\text{e}}^{-1}} \mathbf S {\hat{\mathbf F}_{\text{e}}^{-\text{T}}}.
\end{equation}
Implementing this stress measure, hyperelastic relations are assumed in the form of
\begin{equation}\label{FreeEnerg}
\hat{\mathbf S} = 2 \rho_{\scriptscriptstyle \text{R}}
\frac{\displaystyle \partial \psi(\hat{\mathbf{C}}_{\text{e}})}
{\displaystyle \partial \hat{\mathbf{C}}_{\text{e}}},
\end{equation}
where $\hat{\mathbf C}_{\text{e}}  := \hat{\mathbf F}_{\text{e}}^{\text{T}} \hat{\mathbf F}_{\text{e}}$ stands for the elastic right Cauchy-Green tensor,
$\psi(\hat{\mathbf{C}}_{\text{e}})$ is the free energy per unit mass and $\rho_{\scriptscriptstyle \text{R}}$
denotes the mass density in the reference configuration.
In what follows we suppose that the elastic potential $\psi(\hat{\mathbf{C}}_{\text{e}})$
is isotropic.
Next, the plastic strain rate $\hat{\mathbf{D}}_{\text{p}}$ is introduced:
\begin{equation}\label{plastStrRate}
\hat{\mathbf{D}}_{\text{p}} := \text{sym}(\dot{\mathbf{F}}_{\text{p}} \mathbf{F}^{-1}_{\text{p}}).
\end{equation}
The yield function is formulated using the Kirchhoff stress tensor $\mathbf S$
\begin{equation}\label{YieldCondMiehe}
f := \| \mathbf{S}^{\text{D}} \| - \sqrt{2/3} K,
\end{equation}
just as was done within the additive hypoelasto-plasticity (cf. Section 3).
Note that this yield function can be formulated in terms of
the Mandel tensor, operating on the reference configuration:
\begin{equation}\label{InvarOverstress}
f = \sqrt{ \text{tr}[ ((\mathbf C \widetilde{\mathbf T})^{\text{D}})^2 ]} - \sqrt{2/3} K.
\end{equation}
A six-dimensional flow rule is postulated on the intermediate configuration adopting the symmetric Mandel tensor
$\hat{\mathbf{C}}_{\text{e}}  \hat{\mathbf S}$
\begin{equation}\label{FlowRuleIntermed}
\hat{\mathbf{D}}_{\text{p}} = \lambda_{\text{p}}
(\hat{\mathbf{C}}_{\text{e}}  \hat{\mathbf S})^{\text{D}},
\end{equation}
in combination with the Kuhn-Tucker constraints
\begin{equation}\label{FlowRuleIntermedKuhnTuck}
f \leq 0, \  \lambda_{\text{p}} \geq 0, \  f \lambda_{\text{p}} =0.
\end{equation}
In order to simplify the numerical treatment,
the constitutive equations can be transformed into
the reference configuration (cf. \cite{Lion1997}).
Introducing ${\mathbf C}_{\text{p}}:=\mathbf F^{\text{T}}_{\text{p}} \mathbf F_{\text{p}}$, the
strain energy function takes the form
\begin{equation}\label{TransformEnergy}
\psi(\hat{\mathbf{C}}_{\text{e}}) = \psi(\mathbf C {\mathbf C_{\text{p}}}^{-1}).
\end{equation}
The system of constitutive equations and initial conditions is now
written on the reference configuration as\footnote{In the particular case of Neo-Hookean
strain energy function, an efficient and robust numerical procedure can be developed using
an explicit update formula presented by \cite{ShutovLandgraf}.}
\begin{equation}\label{Simo92Refer}
\widetilde{\mathbf T}  = 2 \rho_{\scriptscriptstyle \text{R}}
\frac{\displaystyle \partial \psi(\mathbf C {\mathbf C_{\text{p}}}^{-1})}
{\displaystyle \partial \mathbf{C}}\big|_{\mathbf C_{\text{p}} =
const}, \ \dot{\mathbf C}_{\text{p}} =  2 \lambda_{\text{p}}
\big( \mathbf C \widetilde{\mathbf T} \big)^{\text{D}} \mathbf C_{\text{p}}, \
\mathbf C_{\text{p}}|_{t=t_0} = \mathbf C^0_{\text{p}}.
\end{equation}
The constitutive relations \eqref{InvarOverstress}, \eqref{FlowRuleIntermedKuhnTuck}, \eqref{Simo92Refer} \emph{are w-invariant}
under arbitrary isochoric reference change.

\textbf{Proof:}
In order to prove the w-invariance, we modify the
proof presented in Section 1 (cf. equations \eqref{NewPlastStrain}--\eqref{FormalDiffere}).
Let $\textbf{C}(t)$ be a given local loading process.
Consider arbitrary $\textbf{F}_0 = const$,
such that $\det(\textbf{F}_0)=1$.
Since the reference change \eqref{ShiftNonlinear} is
volume preserving, the referential mass density remains invariant
\begin{equation}\label{InvarMassDens}
\rho^{\text{new}}_{\scriptscriptstyle \text{R}} = \rho_{\scriptscriptstyle \text{R}}.
\end{equation}
For the analysed material model,
the set of initial data which appears in \eqref{GenerDefinit3} is
given by $\mathcal{Z}_0 = \{ {\mathbf C}_{\text{p}}^0 \}$.
The initial conditions $\eqref{Simo92Refer}_3$ are transformed according to
\begin{equation}\label{NewInitCond}
{\mathbf C}^{\text{new}}_{\text{p}}|_{t=t_0} = {\mathbf C}^{\text{new} \ 0}_{\text{p}}, \quad
{\mathbf C}^{\text{new} \ 0}_{\text{p}} : =
\mathbf F_0^{-\text{T}} \ \mathbf C^0_{\text{p}} \ \mathbf F_0^{-1}.
\end{equation}
We need to check that the weak invariance relation
\eqref{GenerDefinit4} holds true, where the new loading is now given by
\begin{equation}\label{Newloading}
\mathbf{C}^{\text{new}} (t) := \mathbf{F}_0^{-\text{T}} \mathbf{C} (t) \mathbf{F}_0^{-1}.
\end{equation}
First, we introduce the candidate function
\begin{equation}\label{Newvariable}
{\mathbf C}^{\text{new}}_{\text{p}}(t) := \mathbf{F}_0^{-\text{T}} \mathbf{C}_{\text{p}} (t) \mathbf{F}_0^{-1}.
\end{equation}
Obviously, the initial conditions \eqref{NewInitCond} are satisfied by this candidate.
In the following we will show that this candidate function represents the solution for
the new loading \eqref{Newloading}.
Toward that end, we note that
the potential relation $\eqref{Simo92Refer}_1$ implies the following representation of the Mandel stress tensor
on the reference configuration
\begin{equation}\label{representation}
\mathbf C \widetilde{\mathbf T} = g( \mathbf C {\mathbf C}^{-1}_{\text{p}}),
\end{equation}
where $g$ is a suitable isotropic function.
Relying on functions $\mathbf{C}^{\text{new}}(t)$ and $\mathbf{C}^{\text{new}}_{\text{p}}(t)$ we introduce formally the quantity
$\widetilde{\mathbf T}^{\text{new}}(t)$
\begin{equation}\label{FormalNewStress}
\widetilde{\mathbf T}^{\text{new}}  = 2 \rho_{\scriptscriptstyle \text{R}}
\frac{\displaystyle \partial \psi(\mathbf C^{\text{new}} ({\mathbf C}^{\text{new}}_{\text{p}})^{-1})}
{\displaystyle \partial \mathbf{C}^{\text{new}}}\big|_{\mathbf C^{\text{new}}_{\text{p}}=const}.
\end{equation}
In analogy to \eqref{representation}, we have now
\begin{equation}\label{representationNew}
\mathbf C^{\text{new}} \widetilde{\mathbf T}^{\text{new}} = g( \mathbf C^{\text{new}} ({\mathbf C}^{\text{new}}_{\text{p}})^{-1}).
\end{equation}
Since $g$ is isotropic, $\mathbf{F}_0^{-\text{T}} g( \mathbf C {\mathbf C}^{-1}_{\text{p}})  \mathbf{F}_0^{\text{T}} =
g(\mathbf{F}_0^{-\text{T}}  \mathbf C {\mathbf C}^{-1}_{\text{p}} \mathbf{F}_0^{\text{T}})$.
Combining this with \eqref{Newloading}, \eqref{Newvariable}, and \eqref{representationNew}, for the deviatoric part of the Mandel stress
we arrive at
\begin{equation}\label{Important}
\mathbf{F}_0^{-\text{T}}  (\mathbf C \widetilde{\mathbf T})^{\text{D}} \mathbf{F}_0^{\text{T}} =
(\mathbf C^{\text{new}} \widetilde{\mathbf T}^{\text{new}})^{\text{D}}.
\end{equation}
Combining this with \eqref{InvarOverstress} it can be shown that
the yield function remains invariant
\begin{equation}\label{InvarOverstress2}
f^{\text{new}} := \sqrt{ \text{tr}[ ((\mathbf C^{\text{new}} \widetilde{\mathbf T}^{\text{new}})^{\text{D}})^2 ]} - \sqrt{2/3} K,
\quad f^{\text{new}} (t) \stackrel{\eqref{Important}}{=} f(t).
\end{equation}
Thus, the Kuhn-Tucker conditions \eqref{FlowRuleIntermedKuhnTuck} are identically satisfied if we put
\begin{equation}\label{NewMult}
\lambda^{\text{new}}_{\text{p}}(t) := \lambda_{\text{p}}(t).
\end{equation}
Now, differentiating \eqref{Newvariable} with respect to $t$, we have:
\begin{equation}\label{NewvarDerivative}
\dot{{\mathbf C}}^{\text{new}}_{\text{p}}(t) = \mathbf{F}_0^{-\text{T}} \dot{\mathbf{C}}_{\text{p}} (t) \mathbf{F}_0^{-1}
\stackrel{\eqref{Simo92Refer}_2}{=}
2 \lambda_{\text{p}} \mathbf{F}_0^{-\text{T}}  (\mathbf C \widetilde{\mathbf T})^{\text{D}}  \mathbf{F}_0^{\text{T}}   \mathbf{F}_0^{-\text{T}}
\mathbf C_{\text{p}}  \mathbf{F}_0^{-1}.
\end{equation}
Substituting \eqref{Important}, \eqref{Newvariable}, and \eqref{NewMult} into this relation, we obtain
\begin{equation}\label{NewvarDerivative2}
\dot{{\mathbf C}}^{\text{new}}_{\text{p}}(t) =
2 \lambda^{\text{new}}_{\text{p}}
\big( \mathbf C^{\text{new}} \widetilde{\mathbf T}^{\text{new}} \big)^{\text{D}} \mathbf C^{\text{new}}_{\text{p}}.
\end{equation}
Thus, the candidate function ${\mathbf C}^{\text{new}}_{\text{p}}(t)$ is indeed a
solution of the problem \eqref{InvarOverstress}, \eqref{FlowRuleIntermedKuhnTuck}, \eqref{Simo92Refer}
where all ``old '' quantities are formally replaced by their new counterparts.
It follows from \eqref{Important} that
\begin{equation}\label{Important2}
\mathbf{F}_0^{-\text{T}}  (\mathbf C \widetilde{\mathbf T}) \mathbf{F}_0^{\text{T}} =
\mathbf C^{\text{new}} \widetilde{\mathbf T}^{\text{new}}.
\end{equation}
After some simple computations, one gets from \eqref{Important2}
the desired weak invariance relation (cf. \eqref{GenerDefinit4})
\begin{equation}\label{GenerDefConcrete}
\widetilde{\mathbf{T}}  =
\mathbf{F}_0^{-1} \widetilde{\mathbf{T}}^{\text{new}}
\mathbf{F}_0^{-\text{T}}.
\end{equation}
Thus, the system \eqref{InvarOverstress}, \eqref{FlowRuleIntermedKuhnTuck}, \eqref{Simo92Refer} is w-invariant under arbitrary
isochoric reference change.  $\blacksquare$

Due to the w-invariance, there is no preferred reference configuration for this model:
any reference can be used if the mass density is preserved (cf. \eqref{InvarMassDens}). An arbitrary volume-preserving
change of the reference implies merely a certain transformation
of initial conditions according to \eqref{NewInitCond}. Thus, at the stage of constitutive modeling,
there is no need to specify the reference explicitly.

\textbf{Remark 6.}
Let us consider an isotropic hyperelastic material in the popular form
\begin{equation}\label{ClassHyper}
\widetilde{\mathbf T}  = 2 \rho_{\scriptscriptstyle \text{R}}
\frac{\displaystyle \partial \psi(\mathbf C )}
{\displaystyle \partial \mathbf{C}}.
\end{equation}
Roughly speaking, this type of material is covered by the model of \cite{Simo92} as a special case.
Indeed, for the large yield stress (as $K \rightarrow \infty$) the
plastic flow is frozen ($\mathbf C_{\text{p}} = const$) and $\eqref{Simo92Refer}_1$
is reduced to \eqref{ClassHyper} by eliminating $\mathbf C_{\text{p}}$.
Observe that the hyperelastic relation \eqref{ClassHyper}
is w-invariant under arbitrary isochoric reference change only if
it describes an elastic fluid, namely
\begin{equation}\label{ElFluid}
\psi(\mathbf C ) = \psi_{\text{vol}}(\det \mathbf C ).
\end{equation}
On the other hand, the model of \cite{Simo92} is w-invariant even for general (isotropic) free energy $\psi$.
This seeming contradiction is resolved by recognizing that a w-invariant material model of hyperelasticity
is given by $\eqref{Simo92Refer}_1$ and not by \eqref{ClassHyper}.
In contrast to \eqref{ClassHyper}, the w-invariant relation $\eqref{Simo92Refer}_1$
includes a tensor-valued internal constant $\mathbf C_{\text{p}}$.

A different performance of relations $\eqref{Simo92Refer}_1$ and \eqref{ClassHyper} regarding
the w-invariance reveals the following fundamental difference between
these two types of hyperelasticity.
In the case of relation \eqref{ClassHyper}, the function $\psi$ contains information about the stress-free configuration
and this configuration must be known a priori.
No information about the stress-free configuration is contained in $\psi$ which appears in $\eqref{Simo92Refer}_1$.
Since the stress-free configuration does not
have to be known during construction of the free energy function, the
relations of $\eqref{Simo92Refer}_1$ are useful for applications dealing with pre-stressed hyperelastic structures.
$\blacksquare$

\textbf{Remark 7.}
For the material model under consideration, it follows from \eqref{Newloading} and \eqref{Newvariable} that
\begin{equation}\label{Covariance}
\psi(\mathbf C {\mathbf C_{\text{p}}}^{-1}) = \psi(\mathbf C^{\text{new}} {\mathbf C^{\text{new}}_{\text{p}}}^{-1}).
\end{equation}
If $\mathbf C_{\text{p}}$ is understood as a certain referential metric tensor (which is not
necessary related to plastic strain), then the invariance property \eqref{Covariance}
corresponds to the so-called material covariance of the strain energy (cf. Section 3.4 in \cite{MarsdenHughes} or
equation (3.9) in \cite{LuPapado2004}).
$\blacksquare$

\textbf{Remark 8.}
Note that the simplifying assumptions of isotropic elasticity \eqref{FreeEnerg} and
isotropic yield function \eqref{YieldCondMiehe} made in
this section are not crucial for the w-invariance. The w-invariance can be shown even for
a broad class of anisotropic material models, based on a double split of the deformation gradient \citep{Lion2000, Helm,
Tsakmakis2004, Dettmer2004, Henann}.
In particular, dealing with a model of multiplicative viscoplasticity with
nonlinear kinematic hardening \citep{Shutov1}, the w-invariance was proved by
\cite{ShutovPfeiffer}.

It would be wrong to assume that all
plasticity models with the multiplicative split \eqref{Simo1} are automatically w-invariant.
In particular, some models of additive hypoelasto-plasticity with the logarithmic rate
can be rendered in the multiplicative format \citep{Xiao2000IJP}. As we already know
from Section 3, such models are not w-invariant. $\blacksquare$

\section{Discussion and conclusion}

Material properties such as ``isotropic'',
``fluid'', ``solid'' etc. are defined by certain (strong) invariance requirements imposed upon
the constitutive relations \citep{TruesdNoll}.
In this work, a new notion of weak invariance under the reference change is introduced, which
involves a proper transformation of the initial conditions and leaves the remaining constitutive
equations unchanged.
Just like the material symmetries are characterized
by the strong invariance under a certain group of transformations of the reference,
the notion of w-invariance yields new material symmetries in a broader sense.
Especial usefulness of the w-invariance lies in its ``weakness'' since it can
be imposed on solids even under general isochoric transformations.
A meaningful classification of elasto-plastic models is obtained
using this notion. In particular, the w-invariance can be implemented as a tool
for a systematic study of different approaches to finite strain
elasto-plasticity.

Obviously, similar ideas can be implemented within
viscopalsticity and viscoelasticity.
The concept of w-invariance should be compatible with any internal constraints imposed on the material.
In the current study, the incompressibility condition $\det (\mathbf{F}_0) =1$
stems naturally from the incompressibility of the plastic flow. In that way,
the considerations are restricted to the group of unimodular transformations.\footnote{
On the other hand, for some applications,
the incompressibility assumption can be dropped. For instance, a purely volumetric
reference change with isotropic $\mathbf{F}_0$ was considered by \cite{Landgraf}.}
For materials with additional internal constraints, a proper subgroup
of the unimodular group can be considered.

No preferred reference configuration exists for weakly invariant
material models. If a certain model does not fulfill the
weak invariance, the concrete choice of the reference
configuration should be explicitly specified at the stage of
constitutive modeling in order
to avoid ambiguity.

Apart from these theoretical aspects, there is a clear impact of the
w-invariance on the practical applicability of material models.
The w-invariance is useful for
the simulation of multi-stage forming processes,
since each intermediate state can be considered as a new reference
configuration (cf. \cite{ShutovPfeiffer}).
Another important aspect is that the deformation prehistory
of the as-received metallic material is often unknown.
As a result, the as-received state is typically associated with a new reference.
Such lax treatment of the reference configuration is justified for w-invariant models.

The notion of the w-invariance can be used to simplify the mathematical analysis
of constitutive equations (cf. \cite{ShutovKr2010, ShutovPanh}).
Furthermore, any time discretization technique should preserve the
material symmetries of the constitutive relations. For the same reasons, the numerical methods
should retain the w-invariance since it represents a generalized material symmetry.
Some advantages of the numerical algorithms which exactly preserve the w-invariance
were discussed by \cite{ShutovLandgraf} in the specific context of multiplicative inelasticity.

The w-invariance of constitutive equations
should not be confused with the concept of material isomorphism \citep{Bertram1998, Bertram2003}.
The later represents an axiomatic approach to material modeling, based on the assumption that the
elastic properties remain constant. The w-invariance is a generalized symmetry restriction
which does not necessarily imply constant elastic properties.

Note that only some basic approaches to elasto-plasticity
were tested in this study using the simple case
of elastic-perfectly-plastic material.
This test is sufficient as it includes the relevant
effects of the geometric nonlinearity.

In the field of additive hypoelasto-plasticity, two basic restrictions are commonly adopted:
the consistency criterion of Prager (1st consistency condition) and the need to avoid any energy dissipation
in the elastic range (2nd consistency condition). Recall that the constitutive
relations based on the logarithmic stress rate are the only option if
both consistency conditions have to be satisfied (the case of constant elastic
moduli is considered here). Unfortunately, as it was shown in Section 3,
models based on the logarithmic rate are not w-invariant.
Thus, within the additive hypoelasto-plasticity, it is impossible to combine the mentioned consistency conditions
and the w-invariance under arbitrary isochoric reference change.
Another modeling framework relies
on geometrically linear equations of
elasto-plasticity in combination with the Lagrangian Hencky strain.
An essential advantage of this framework lies in its simplicity.
This setting is free from the non-physical energy dissipation
in the case of frozen plastic flow. Unfortunately, as shown in Section 4, the material model is not w-invariant.
Finally, it was shown in Section 5 that the framework of multiplicative hyperelasto-plasticity can be formulated in a
consistent manner such that any isochoric change of the reference configuration
would merely imply a transformation of the initial conditions. Table \ref{tab1} is convenient to summarize
the main results of the current study.

\begin{table}[h]
\caption{Summary of the fundamental properties for some
basic models of finite strain elasto-plasticity.}
\begin{tabular}{| l | | l | l | l |}
\hline
Type of model  & Energy dissipation   &  Shear   &  w-invariance under    \\
               & in elastic range  &   oscillation  &  reference change   \\ \hline \hline
Additive hypoelasto-plasticity:   &        &            &    \\ \hline
Zaremba-Jaumann rate   & dissipation      &  oscillation         & w-invariant   \\ \hline
Green-Naghdi rate   & dissipation        & no oscillation        & not w-invariant   \\ \hline
Logarithmic rate   & no dissipation        & no oscillation        & not w-invariant   \\
                   & \footnotesize{(for constant elastic moduli)}        &          &    \\
\hline \hline
Add. plast. with log. strain   &  no dissipation        & no oscillation        & not w-invariant   \\ \hline \hline
Multiplicative plasticity   &  no dissipation        & no oscillation        & w-invariant   \\ \hline
\end{tabular}
\label{tab1}
\end{table}

\section*{Acknowledgement}
The authors are grateful to the reviewers for their helpful suggestions on the manuscript.
The financial support provided by DFG within SFB 692 is likewise acknowledged.

\section*{Appendix A}
Let $b_1$, $b_2$, $b_3$ be the eigenvalues of  $\mathbf{B}$.
We recall the closed form coordinate-free expression for the log-spin $\mathbf{\Omega}^{\text{log}}$
(see, for instance, \cite{XiaoBruhns1997})
\begin{equation*}\label{LogSpinExplicitF}
\mathbf{\Omega}^{\text{log}} = \mathbf{W} + \mathbf{N}^{\text{log}},
\end{equation*}
\begin{equation*}\label{LogSpinExplicitF2}
\mathbf{N}^{\text{log}} = \begin{cases}
    \mathbf{0} \quad \quad \quad \quad \quad \quad \quad \quad \quad \quad \quad \quad \quad \quad \quad \quad \quad  \text{if} \ b_1=b_2=b_3 \\
    2 \ \text{skew}(\nu \ \mathbf{B} \mathbf{D}) \quad \quad \quad \quad \quad \quad \quad \quad \quad \quad \quad \ \text{if} \ b_1 \neq b_2=b_3 \\
     2 \ \text{skew}(\nu_1 \ \mathbf{B} \mathbf{D} + \nu_2 \ \mathbf{B}^2 \mathbf{D} +
     \nu_3 \ \mathbf{B}^2 \mathbf{D} \mathbf{B}) \quad \text{if} \ b_1 \neq b_2 \neq  b_3 \neq b_1
\end{cases},
\end{equation*}
\begin{equation*}\label{LogSpinExplicitF3}
\nu = \frac{1}{b_1-b_2} \Big( \frac{1 + b_1/b_2}{1 - b_1/b_2} + \frac{2}{\ln(b_1/b_2) }\Big),
\end{equation*}
\begin{equation*}\label{LogSpinExplicitF4}
\nu_j = \frac{1}{\Delta } \sum_{i=1}^3 (-b_i)^{3-j}  \Big( \frac{1 + \varepsilon_i}{1 - \varepsilon_i} + \frac{2}{\ln (\varepsilon_i)}  \Big), \ j =1,2,3,
\end{equation*}
\begin{equation*}\label{LogSpinExplicitF5}
\Delta = (b_1-b_2)(b_2-b_3)(b_3-b_1), \
\varepsilon_1 = b_2/b_3, \ \varepsilon_2 = b_3/b_1, \ \varepsilon_3 = b_1/b_2.
\end{equation*}

\bibliographystyle{elsarticle-harv}

\

Corresponding author: \\ Alexey Shutov. \\
e-mail: \texttt{alexey\_shutov79@mail.ru} \\
web: http://sites.google.com/site/materialmodeling

\end{document}